\documentclass[conference]{IEEEtran}
\usepackage{url} 
\usepackage{amsmath, amssymb, amsthm}
\usepackage{hyperref}
\usepackage{graphicx}
\usepackage{gensymb} 
\usepackage[caption=false]{subfig}

\usepackage{float}
\begin{document}

\title{The Economic Dispatch for Integrated Wind Power Systems Using Particle Swarm Optimization}

\author{\IEEEauthorblockN{Mohamed Abuella}
\IEEEauthorblockA{Departmtent of Electrical and Computer Engineering\\
University of North Carolina at Charlotte\\
Charlotte, USA\\
Email: mabuella@uncc.edu}
\and
\IEEEauthorblockN{Constantine J. Hatziadoniu}
\IEEEauthorblockA{Department of Electrical and Computer Engineering\\
Southern Illinois University\\
Carbondale, USA\\
Email: hatz@siu.edu }}

\maketitle

\begin{abstract}
The economic dispatch of wind power units is quite different from that in conventional thermal units; since the adopted model should take into consideration the intermittency nature of wind speed as well. Therefore, this paper uses a model that takes into account the aforementioned consideration in addition to whether the utility owns wind turbines or not. The economic dispatch is solved by using one of the modern optimization algorithms: the particle swarm optimization algorithm. A 6-bus system is used and it includes wind-powered generators besides to thermal generators. The thorough analysis of the results is also provided.\\

\indent{{\textbf{\textit{Keywords|Economic dispatch; particle swarm;  wind energy.}}}}\\
\end{abstract}

\section{Introduction} \label{intro}
As a result of the intermittency nature of wind power,  the economic dispatch of resources that include wind power is quite different of that in pure conventional thermal units. Therefore, how can the economic dispatch of this promised future resource of energy be achieved? \\
Various mathematical programming approaches used to solve this kind of optimization problem in power systems, based on linear and nonlinear programming were proposed, including Newton method, quadratic programming, and interior-point method \cite{Wang2005}.
The mathematical methods utilize the first or second derivative information in essence. In this way, it is apt to fall into local optima. Furthermore, there is a difficulty of applying gradient-based optimization techniques. Therefore, various non-classical optimization methods have emerged to cope with some of the traditional optimization algorithms' shortcomings.
The main modern optimization techniques are genetic algorithm (GA), evolutionary programming (EP), artificial neural network (ANN), simulated annealing (SA), ant colony optimization (ACO), and particle swarm optimization (PSO). They have been successfully applied to wide range of optimization problems in which global solutions are more preferred than local ones \cite{Vlachogiannis2009},\cite{hwary_enhanced}.\\
Kennedy and Eberhart first introduced particle swarm optimization (PSO) in 1995 as a new heuristic method \cite{Ken}.
In \cite{hwary_enhanced} there is a comprehensive coverage of different PSO applications in solving optimization problems in the area of electric power systems up to 2006.
 The review in \cite{Review2009}  is about the historical research production of the economic dispatch considering the wind power, besides that it also presents some models and different optimization algorithms as well.

 In 2008 \cite{Hetzer} is one of the pioneer studies about the economic dispatch including the wind power was reported. It also includes the definitions about the wind power cost and its factors in wind energy conversion systems (WECS) combining both cases, whether the operator owns WECS or not. In addition to the direct cost of wind power, cost factors of the overestimation and underestimation of wind power have also been proposed.
 
 This paper is intending to investigate the interconnection of wind generators besides the conventional generators into power systems and its impact on the generation resource management.
The rest of this paper is organized as follows.
Section~\ref{problem statement} introduces the motivation and the problem statement.
Section~\ref{Ch_wind} it discusses the analysis and characterization of wind speed and power.
Section~\ref{PSO} the particle swarm optimization (PSO) algorithm is described. 
In Section~\ref{model} the results of the implementation of PSO to find the economic dispatch of a benchmark system are discussed to some extent. Finally, Section~\ref{conclusion} gives the conclusions. 

\section{Problem Formulation}  \label{problem statement}
\subsection{Problem objectives}
The objective function of the optimization problem in this paper is to minimize the operating cost of power generation power from a combination of wind-powered and conventional generators. The operating cost of conventional thermal generators is represented by a quadratic equation as following \cite{Wood}:
\begin{equation}\label{eq cc}
    C_{i}=a_{i}p_{i}^{2}+b_{i}p_{i}+c_{i}
\end{equation}
Where $p_{i}$ is the generation power from the \textbf{\emph{i}}th conventional generator; and \emph{a, b} and \emph{c} are the operating cost coefficients of the \textbf{\emph{i}}th generator.
 The wind power generation cost $C_{w}$ which may be not exist if the power operator owns the wind powered-generators, but it could be considered as a payback cost or  a maintenance cost \cite{Hetzer}:
\begin{equation}\label{eq cw}
    C_{w,i}=d_{i}w_{i}
\end{equation}
Since $w_{i}$ is the scheduled wind power from the \textbf{\emph{i}}th wind-powered generator; and $d_{i}$ is the direct cost coefficient for the {\emph{i}}th wind generator.

Because of the uncertainty of generated wind power, there are two scenarios of wind power costs. 
The surplus of wind power as a result of underestimation of the available wind power and hence scheduling the wind power $w_i$ less than what it would be. Thus, $C_{p}$ appears as a penalty cost \cite{Hetzer}.
\begin{equation}\label{eq cp}
 C_{p,i}=k_{p,i}\int_{w_{i}}^{w_{r,i}}(w-w_{i})f_{w}(w)dw
\end{equation}
Where $f_{w}$ is the Weibull distribution function for wind power, for more details see section (\ref{sec iowecs}); and $k_{p}$ is the penalty cost coefficient.

On the other hand, a deficit of wind power which occurs by the overestimation of the available wind power and scheduling the wind power $w_i$ more than it would be available. At that situation, the deficit will be compensated by a reserve power sources. That means there is also a cost for the deficit of wind power.
Thus, $C_{r}$ is presented as a reserve cost.
\begin{equation}\label{eq cr}
 C_{r,i}=k_{r,i}\int_0^{w_{i}}(w_{i}-w)f_{w}(w)dw
\end{equation}
$ k_{r,i} $ is the  reserve cost coefficient for the \textbf{\emph{i}}th wind generator.
\subsection{Problem Constraints}
Due to the physical or operational limits in practical systems, there is a set of constraints that should be satisfied throughout the system operations for a feasible solution \cite{Wang1}.
\begin{itemize}
  \item{Generation capacity constraints}:\\
For normal system operations, real power output of each generator is restricted by lower and upper limits as follows:
\begin{equation}\label{eq pcons}
  p_i^{min}\leq p_{i}\leq p_i^{max}
\end{equation}
\begin{equation}\label{eq wcons}
0\leq w_{i}\leq w_{r,i}
\end{equation}
Since \emph{$ w_{r,i}$} is the rating wind power from the \textbf{\emph{i}}th wind-powered generator.
 \item{The transmission line losses constraints}:\\
\begin{equation} \label{eq losses}
S_{line,i}\leq S_{line,i}^{max}  
\end{equation}
 $S_{line,i}$ is losses of the \textbf{\emph{i}}th transmission line.
  \item{Power balance constraint}:\\
The total power from conventional and wind generators must cover the total demand.
\begin{equation}\label{eq balcons}
\sum_{i=1}^{M} p_i+\sum_{i=1}^{N} w_i=D
\end{equation}
Where \emph{M} number of conventional power generators; \emph{N} number of wind-powered generators; and \emph{D} is the demand which equals to the system load and losses.
\end{itemize}

\subsection{Problem Statement}
In summary, the objectives of optimal economical dispatch is to minimize the operating cost from the conventional and wind-powered generators includes the penalty of underestimation and overestimation of wind power, subject to the certain constraints.

The model of economic dispatch for thermal and wind-powered generators \cite{Hetzer}:
\begin{equation}\label{eq model}
\sum_{i=1}^{M} C_i(p_i)+\sum_{i=1}^{N} C_{w,i}(w_i)+\sum_{i=1}^{N} C_{p,i}(w_{i})+\sum_{i=1}^{N}C_{r,i}(w_{i})
\end{equation}
subject to: The constraints that are represented as in equations (\ref{eq pcons}) - (\ref{eq balcons}).

Note using a classic economic dispatch approach for the model in equation (\ref{eq model}), which takes the partial derivative of the objective function respect to generator outputs; it's difficult due to the integrals in the wind power cost terms as in equations (\ref{eq cp}) and (\ref{eq cr}), Therefore, Particle Swarm Optimization (PSO) algorithm is used for solving this optimization problem.

\section{The Analysis of Wind Speed and Power}  \label{Ch_wind}
\subsection{Probability Analysis of Wind Power}
Before starting the discussion of economic dispatch of systems that contain wind-powered generators, it will be a good idea to identify the wind speed characterization by probability principles and its subsequent transformation to wind power.
\subsection{Wind Speed Characterization}
\begin{figure}[!ht]
\begin{center}
\includegraphics[width=2.25in]{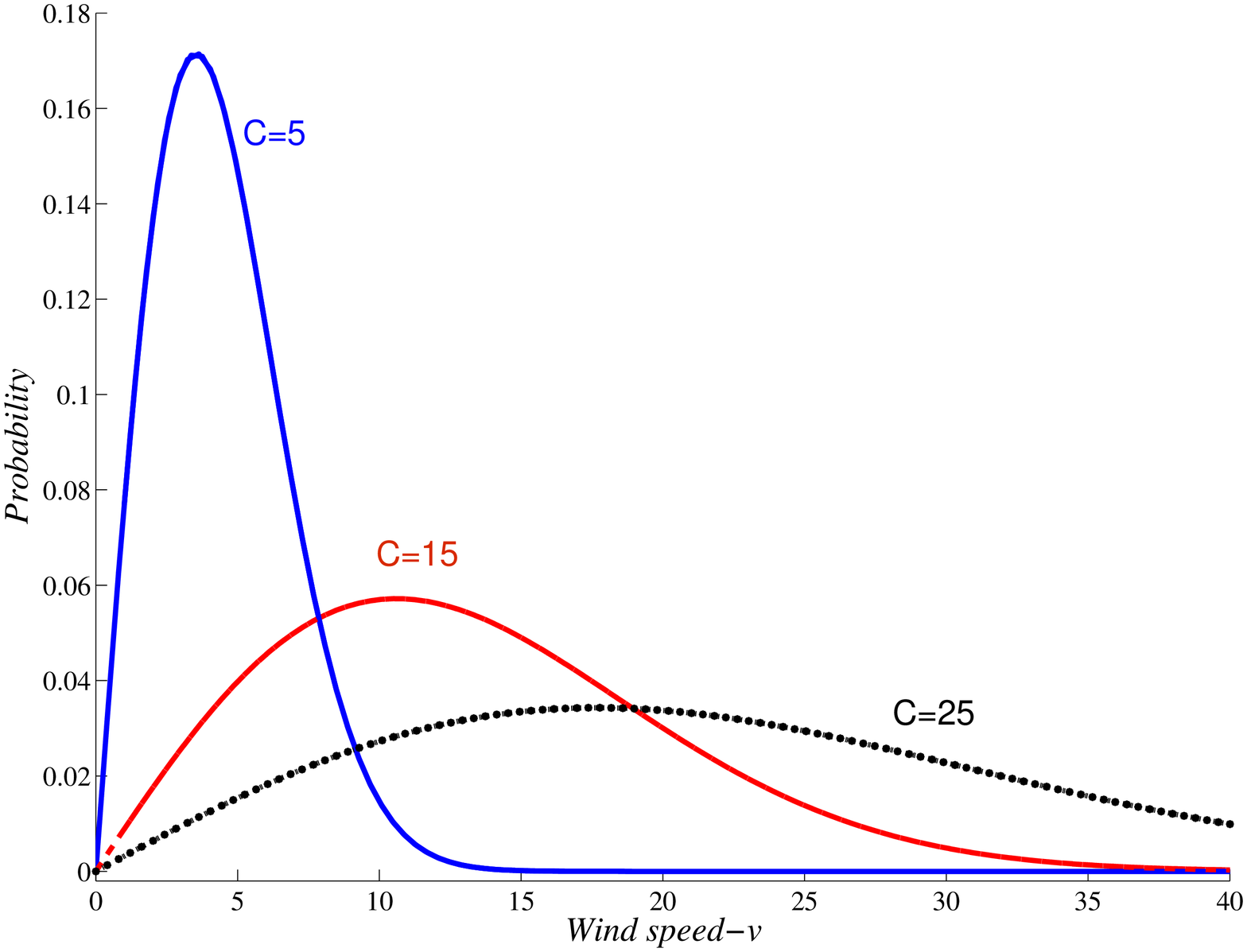}
\caption[Weibull pdf of wind speed for several values of scale factor c]{Weibull pdf of wind speed for several values of scale factor c \cite{wind&solar} \label{fig:weibull}}
\end{center}
\end{figure}

The wind speeds in a particular place can be considered as a Weibull distribution over time \cite{wind&solar}. The probability density function (pdf) of the Weibull distribution $f_{V}(v)$ is given by:
\begin{equation}\label{eq wie}
f_{V}(v)=\left(\frac{k}{c}\right)\left(\frac{v}{c}\right)^{(k-1)}\;e^{-(\frac{v}{c})^k},\qquad  0<v<\infty
\end{equation}
Where \emph{v} is the wind speed; \emph{c} is scale factor; \emph{k} is the shape factor.
Fig. \ref{fig:weibull} illustrates the Weibull pdf with shape factors \emph{k}=2, and curves of scale factor $c=$ 5 $m/s$, 15 $m/s$, and 25 $m/s$ are indicated.\\
\\
The cumulative distribution function (cdf) of Weibulll distribution $F_{V}(v)$ is obtained by integration of pdf:
\begin{equation}\label{eq wie}
F_{V}(v)=\int_0^{v}f_{V}(v)\,dv=1-e^{-(\frac{v}{c})^k}
\end{equation}

\subsection{WECS Input/Output and Probability Functions}\label{sec iowecs}
\begin{figure}[!ht]
\begin{center}
\includegraphics[width=2.3in]{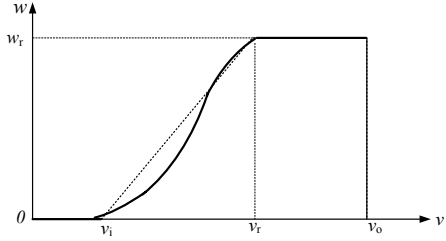}
\caption[The captured wind power curve]{The captured wind power curve \cite{wind&solar} \label{fig:wcurve}}
\end{center}
\end{figure}

For wind energy conversion systems (WECS) as it is shown in Fig. \ref{fig:wcurve}, the wind power curve from probability point of view can be represented in three regions as in equation (\ref{eq linw}) \cite{Hetzer}.
\begin{equation}\label{eq linw}
  w = \left\{
  \begin{array}{l l}
    0; & \quad (v < v_i\;or\; v \geq v_o)\\
    w_r\frac{(v-v_i)}{(v_r-v_i)}; & \quad (v_i \leq v < v_r)\\
    w_r; & \quad (v_r \leq v < v_o)\\
  \end{array} \right.
\end{equation}
Where \emph{w} is the wind power; \emph{$w_r$} is the rating power of WECS; \emph{$v_i$} is the cut-in wind speed; \emph{$v_o$} is the cut-out of wind speed; \emph{$v_r$} is the rating wind speed at which the rating power \emph{$w_r$} is captured.

The linear transformation from wind speed to wind power in the linear region $ (v_i \leq v < v_r)$ is done as following \cite{Hetzer}:
\begin{equation}\label{}
\because w=T(v)=av+b \hspace{0.1in}  \therefore   v=T^{-1}(w)  \Rightarrow   v=\frac{(w-b)}{a}  \nonumber     
\end{equation}
So now $v$ in terms of $w$, and thus $f_{V}(v)$ $\Rightarrow$  $f_W(w)$ as follows:

\begin{equation}\label{eq wtrans}
 \therefore f_W(w)=f_V\left(\frac{w-b}{a}\right)\left|\frac{1}{a}\right|
\end{equation}
where:\\
$T$ is the general transformation;
$w$ wind power random variable;
$v$ wind speed random variable;

For Weibull distribution function, the transformation will lead to discrete and continuous ranges as following:
For discrete portions:
\begin{equation}\label{}
Pr\{W=0\}=F_V(v_i)+(1-F_V(v_o))=1-e^{-(\frac{v_i}{c})^k}+e^{-(\frac{v_o}{c})^k}  \end{equation}
\begin{equation}\label{}
Pr\{W=w_r\}=F_V(v_o)-F_V(v_r)=-e^{-(\frac{v_r}{c})^k}-e^{-(\frac{v_o}{c})^k}
 \end{equation}

 While for the continuous portion of the wind power curve:
 \begin{equation}\label{cont eq}
f_W(w)=\frac{klv_i}{w_r c}\left(\frac{(1+\rho l)v_i}{c}\right)^{(k-1)}e^{-\left(\frac{(1+\rho l)v_i}{c}\right)^k}
 \end{equation}
\begin{equation}\label{}
\rho=\frac{w}{w_r},  \hspace{0.25in} \l=\frac{(v_r-v_i)}{v_i} \nonumber 
 \end{equation}

\section{Particle Swarm Optimization Algorithm}  \label{PSO}
The PSO is originally suggested by Kennedy and Eberhart based on the analogy of swarm of bird and school of fish \cite{Ken}. The algorithm was simplified and used for solving the optimization problems.
\subsection{Standard PSO Algorithm}
\begin{figure}[!ht]
\begin{center}
\includegraphics[width=1.4in]{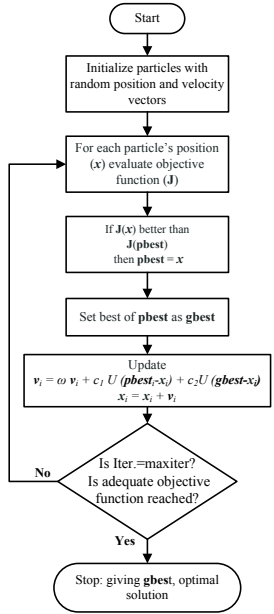}
\caption[PSO algorithm flowchart]{PSO algorithm flowchart\label{fig:psofchart}}
\end{center}
\end{figure}

The following is the conventional terminology of the parameters in PSO:
Let \emph{x} and \emph{v} denote a particle coordinates (position) and its corresponding speed magnitude (velocity) in a search space, respectively. Therefore, the\textbf{ \emph{i}}th particle is represented as $x_i=[x_{i1}, x_{i2},....,x_{im}]$. Since \textbf{\emph{m}} is the last dimension or coordinate of the position of the the\textbf{ \emph{i}}th particle in the search space and so that the dimension $\textbf{\emph{d}}=1,2,...,\textbf{ \emph{m}}$.\\
The best previous position of the\textbf{ \emph{i}}th particle is saved and represented as \cite{Gaing},\\ $pbest_i=[pbest_{i1}, pbest_{i2},....,pbest_{im}]$.\\
The position of the best particle among all the particles in the group is represented by the $gbest$.
In a particular dimension \textbf{\emph{d}} there is a group best position which is $gbest_d$.\\
The velocity for the\textbf{ \emph{i}}th particle is represented as, $v_i=[v_{i1}, v_{i2},....,v_{id}]$.
The modified velocity and the position of each particle can be calculated by using the following formulas:
\begin{equation} \label{eq pso v}
v_{id}^{k+1}=w*v_{id}^{k}+c_1*U*(pbest_{id}^k-x_{id}^k)+c_2*U*(gbest_{d}^k-x_{id}^k)
\end{equation}
\begin{equation} \label{eq pso x}
x_{id}^{k+1}=x_{id}^k+v_{id}^{k+1}
\end{equation}
$i=1,2,....,n; \qquad d=1,2,...,m$\\
Where $x_{id}^k,\! v_{id}^k$ the position and the velocity of the\textbf{ \emph{i}}th particle in the\textbf{ \emph{d}}th dimension at an iteration $k$; $n$ number of particles in a group; $m$ number of members in a particle; $w$ inertia weight factor; $c_1,c_2$ acceleration factors; $U$ uniform random number in the range [0,1];
\\
The velocity should between $v_d^{min} \leq v_{id} \leq v_d^{max}$ If $v_d^{max}$ is too high, particles might move past good solutions. While if $v_d^{max}$ is too small, particles may not explore sufficiently beyond local solutions. In many experiences with PSO, was often set at 10 - 20\% of the dynamic range of the variable on each dimension \cite{Gaing}.\\
The constants $c_1$ and $c_2$ represent the weighting of the stochastic acceleration terms that pull each particle toward the $pbest$ and $gbest$ positions. The acceleration constants $c_1$ and $c_2$ are often set to be 2 according to past experiences \cite{Gaing}.
Suitable selection of inertia weight $w$ in equation (\ref{eq pso v}) provides a balance between global and local explorations, to find a sufficiently optimal solution. In general, the inertia weight is set according to the following equation:
\begin{equation} \label{eq pso x}
w=w_{max}-\frac{(w_{max}-w_{min})}{iter_{max}}\times iter
\end{equation}
Where $iter_{max}$ is the maximum number of iterations (generations). PSO algorithm flowchart is shown in Fig. \ref{fig:psofchart}.

\section{PSO Implementation on IEEE Benchmark System}  \label{model}
 PSO algorithm is coded in MATLAB to determine the economic dispatch of the generation power for an electrical power system with considering various levels of wind power penetration. In this paper, the PSO algorithm is implemented for solving the economic dispatch of a 6-bus system that includes wind-powered generators in addition to the conventional generators. Since the used model in equation (\ref{eq model}) has an objective function that contains some terms that need integration, the PSO algorithm can be applied efficiently to solve this type of objective functions.
\subsection{The Data of The System}
\begin{figure}[!ht]
\begin{center}
\includegraphics[width=2.3in]{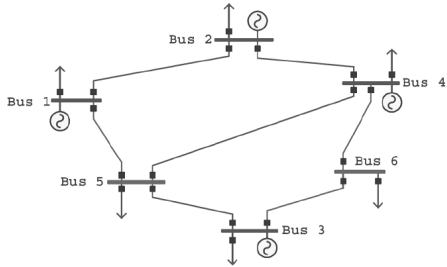}
\caption[Single-line diagram of 6-bus  system]{Single-line diagram of 6-bus system \cite{abuella} \label{fig:6bussystem}}
\end{center}
\end{figure}

The 6-Bus System in Fig. \ref{fig:6bussystem} is adopted to calculate the optimal economic dispatch of  generation power including wind power. This system consists of six buses and four generators at buses 1, 2, 3, and 4, generators at buses 3 and 4 are wind-powered generators. There are seven transmission lines, and there are no LTC transformers or VAR compensation devices in this system. For all detailed data of this 6-bus system refer to \cite{abuella}. 

\begin{table}[!ht]
\caption{Generators data of 6-bus system\label{table:6gendata}}
\begin{center}
\includegraphics[width=2.6in]{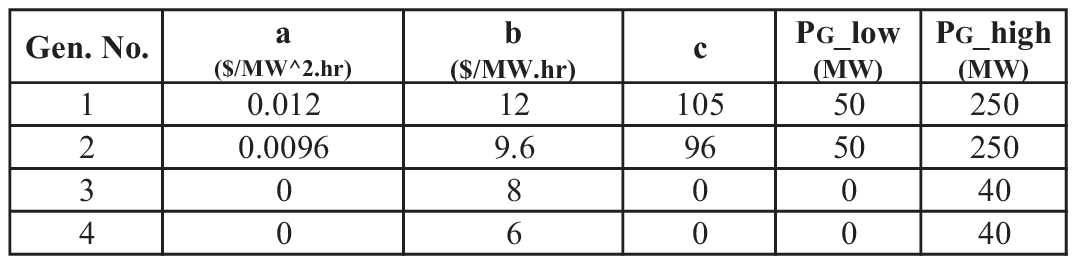}
\end{center}
\end{table}
As TABLE \ref{table:6gendata} is shown, the system has two conventional thermal units and two wind-powered generators.
 The higher output of each wind-powered generator is 40 $MW$. While the direct cost of wind power 8 and 6 $\$/(MW.hr$) for wind-powered generator 3 and 4 respectively. The difference in direct cost of wind power is for including the variety purpose in the study, and to get more options of dispatching.
 The parameters of the wind turbine are cut-in wind speed $v_i=5 m/s$, rating wind speed  $v_r=15 m/s$, and cut-out wind speed $v_o=45 m/s$.
\subsection{The Objective Function}
The aim is to implement the PSO algorithm with the model which is discussed in section \ref{problem statement}. The objective function is presented by equation (\ref{eq model}), the term $\sum_{i=1}^{M} C_i(p_i)$ is the cost of the real power of thermal-generators, $\sum_{i=1}^{N} C_{w,i}(w_i)$ is the direct cost of wind power, $\sum_{i=1}^{N} C_{p,i}(w_{i})$ is the penalty cost of the underestimation of the available wind power, and $\sum_{i=1}^{N}C_{r,i}(w_{i})$ is the reserve cost of the overestimation of the available wind power. The two latter terms in the objective function have integrals as they are represented in equations (\ref{eq cp}) and (\ref{eq cr}).

\begin{figure}[!ht]
\begin{center}
\includegraphics[width=2.3in]{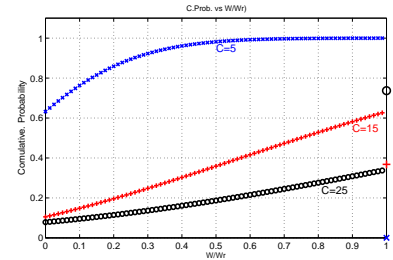}
\caption[Cumulative probability distribution of wind power vs. normalized wind power]{Cumulative probability distribution of wind power vs. normalized wind power \label{fig:C_Propability}}
\end{center}
\end{figure}

Fig. \ref{fig:C_Propability} illustrates the cumulative probability distribution of wind power, it is produced from the integration of equation \ref{cont eq}. This figure has such importance for the next investigations.\\
Note: $w$ is the available wind power, while $w_r$ is the rating wind power.

\subsection{PSO Solution for Base Case}\label{sec:wcbasecase}
The base case with total load of 400 $MW$. The parameters of Weibull distribution of wind speed here are scale factor $c=$5$ m/s$, while the shape factor $k$ is 2. By assumption, the reserve cost coefficient as a result of overestimation of available wind power would be 1 $\$/MW.hr$. On the other hand, penalty cost coefficient as a result of underestimation of available wind power is 0 $\$/MW.hr$, this means the utility owns wind turbines so there is no penalty of surplus produced wind power. The changing of these coefficients and their effect on the total cost will be investigated later.

The economic dispatch of the base case is in TABLE \ref{table:wc400MW}. The minimum cost of real power from both thermal and wind power generators is 4777.49 $\$/hr$.
Wind-powered generators in these conditions supply maximum outputs because they are more economic. While the first thermal generator supplies less power to the system than the second generator because its generated power is more expensive. All generators' cost coffeceitns are in TABLE \ref{table:6gendata}.\\
As it is shown in TABLE \ref{table:wc400MW} the outputs of generators are equal to the demand plus losses in the system.
\begin{table}[!ht]
\caption{PSO result of economic dispatch for base case (400$MW$) \label{table:wc400MW}}
\begin{center}
\includegraphics[width=2.5in]{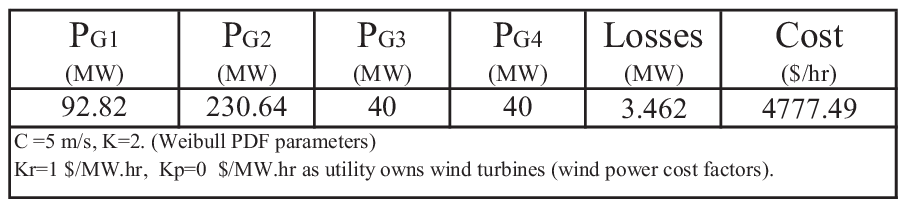}
\end{center}
\end{table}

 \subsection{PSO Solution for Different Loading}
The economic dispatch solution by PSO algorithm when the system load increases gradually is as in TABLE \ref{table:ED_FLcases}. The limits of the transmission lines losses are neglected because they are relatively low.

\begin{table}[!ht]
\caption{PSO result for ED of different load cases of 6-bus system\label{table:ED_FLcases}}
\begin{center}
\includegraphics[width=2.5in]{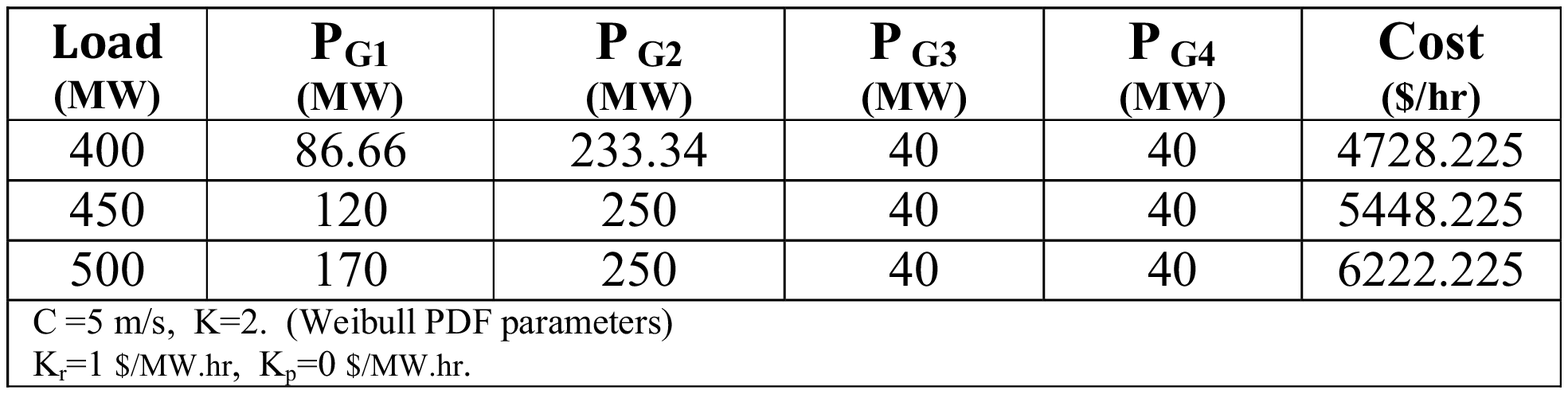}
\end{center}
\end{table}

\subsection{The Effects of Wind Power Cost Coefficients}
Next the variety of wind power cost coefficients and wind speed factors will be investigated. So that their effects on the output schedule of the generators and hence the total cost for the base case 400 $MW$ are presented as following.

The shape factor of wind speed probability Weibull distribution $k=2$ and it is kept constant at this value.
While the scale factor $c$ is changing between 5$m/s$ to 25 $m/s$. The constant direct costs of  wind power from wind-powered generators 3 and 4 are 8 and 6 $\$/(MW.hr)$ respectively.
For sake of convenience, hereinafter the units will be dropped from these coefficients.

\subsubsection{The Effects of Reserve Cost Coefficient}
First, assume that the utility owns wind turbines, so that the penalty cost of additional available wind power over scheduled power will be 0, and this also means the coefficient $k_p=0$ as it is derived from equation (\ref{eq cp}).

\begin{figure}[!ht]
\centering
\subfloat[$k_r$=1]{\label{fig:1kr_outs}\includegraphics[width=0.25\textwidth]{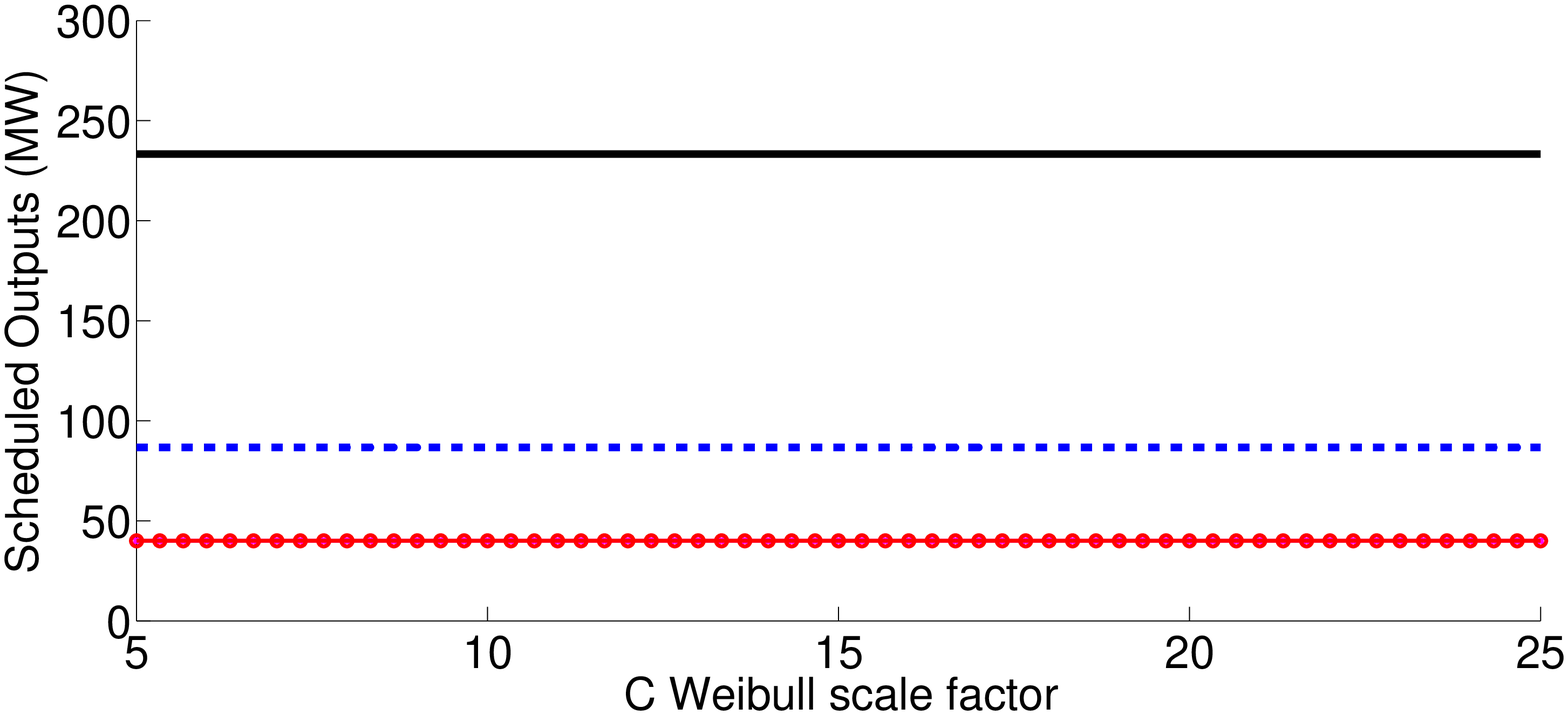}}
\subfloat[$k_r$=10]{\label{fig:10kr_outs}\includegraphics[width=0.25\textwidth]{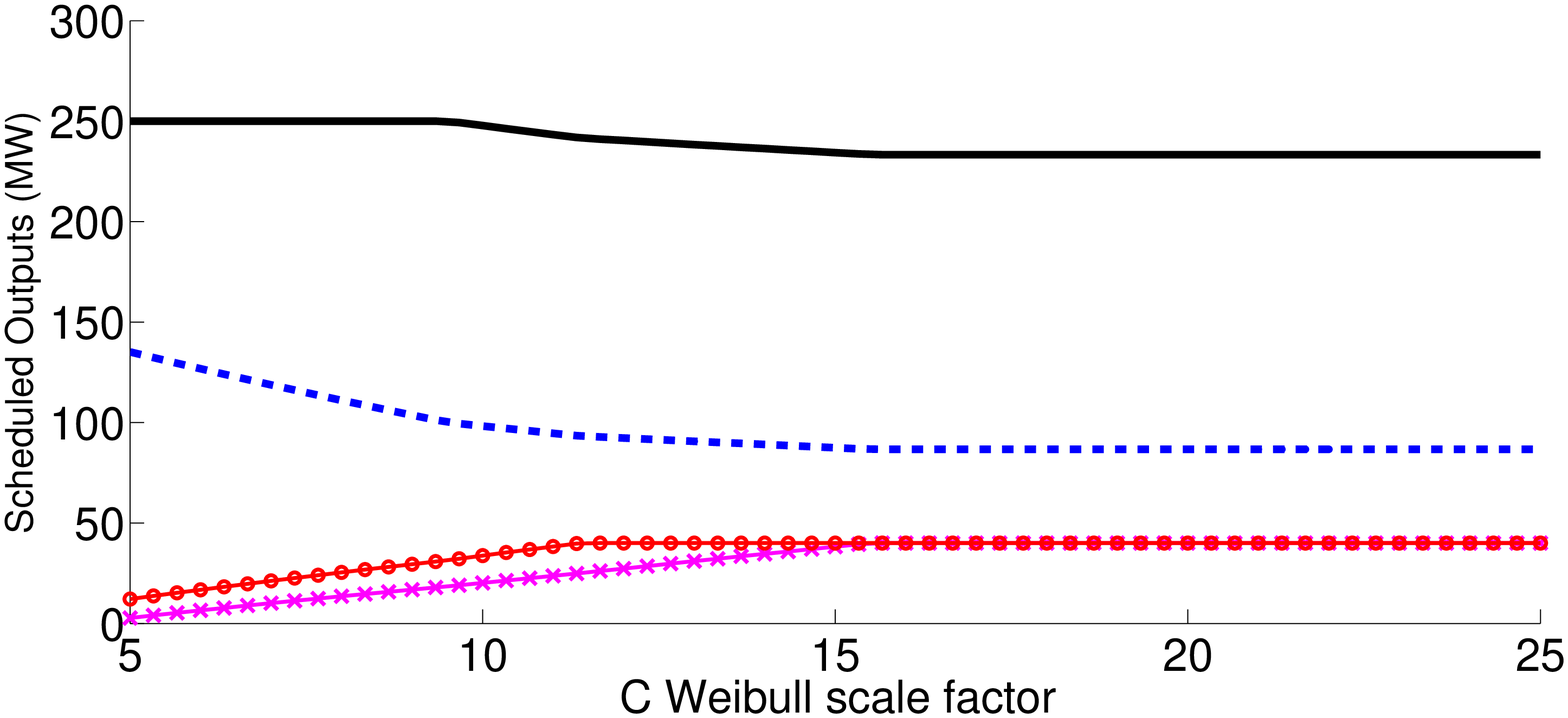}} \\
\subfloat[$k_r$=100]{\label{fig:100kr_outs}\includegraphics[width=0.25\textwidth]{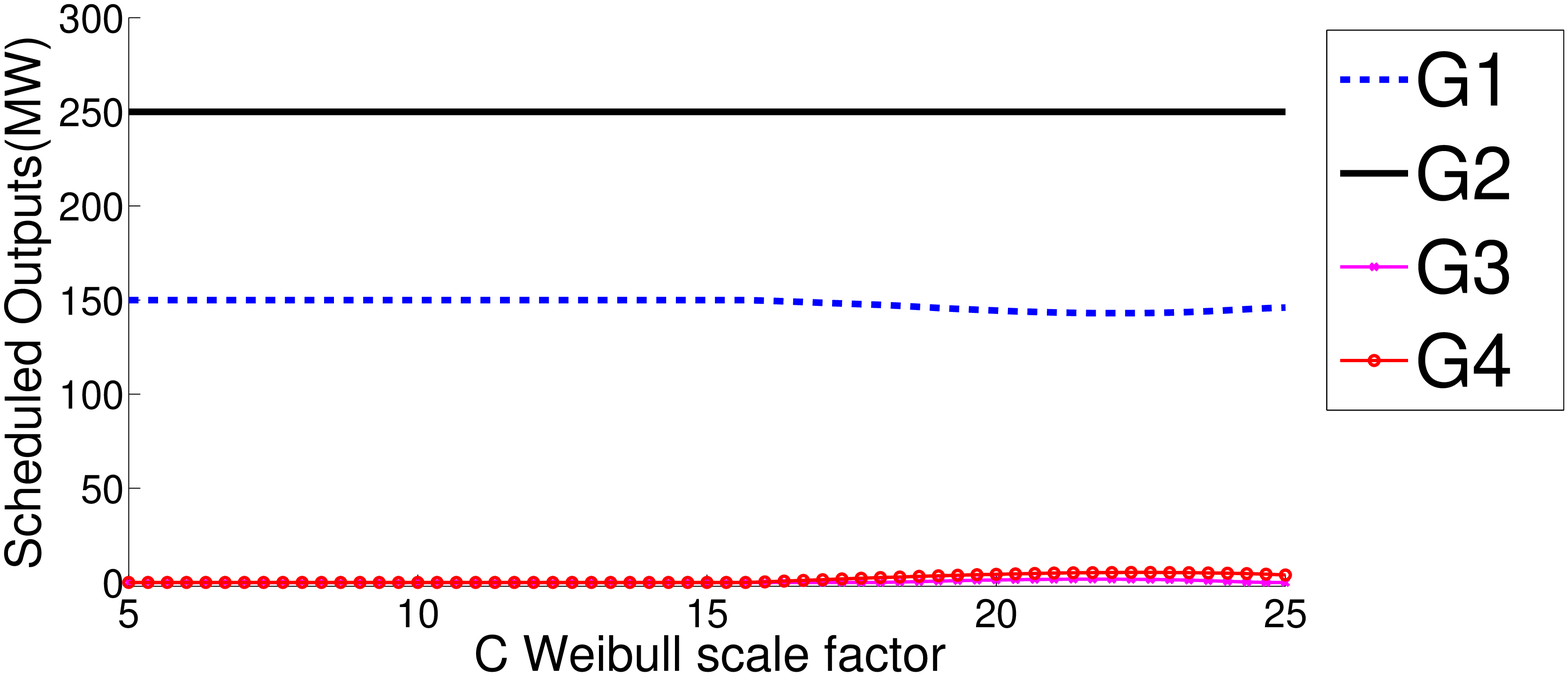}}
\hspace{0.3in}\parbox{3.4in}{\caption[Generator outputs vs. Weibull scale factor (c) for some values of $k_r$,]{Generators' outputs vs. Weibull scale factor (c) for some values of reserve cost in base case 400 $MW$}\label{fig:krsoutputs}}
\end{figure}

Fig. \ref{fig:krsoutputs} shows the result of PSO algorithm for the economic dispatch of generators' outputs as a function of the scale factor of Weibull distribution of wind speed $c$ for different values of reserve cost coefficient $k_r$.

 When c scale factor of Weibull distribution of wind speed increases, the reserve cost decreases. That can be verified from Fig. \ref{fig:C_Propability}. When c increases, the probability of wind power decreases, then the reserve cost reduces as well.
Thus, the reserve cost reduces by increasing c, the scheduled outputs of wind-powered generators will increase gradually as in Fig. \ref{fig:10kr_outs}.
In Fig. \ref{fig:100kr_outs} there is a small increase in wind-powered generators form $c=$20 to $c=$25, it is a small change because the reserve cost coefficient in this case relatively high $k_r$=100.\\

\subsubsection{Critical Reserve Cost Coefficient} 
\begin{figure}[!ht]
\centering
\subfloat[$c$=5]{\label{fig:C5kr_outs}\includegraphics[width=0.25\textwidth]{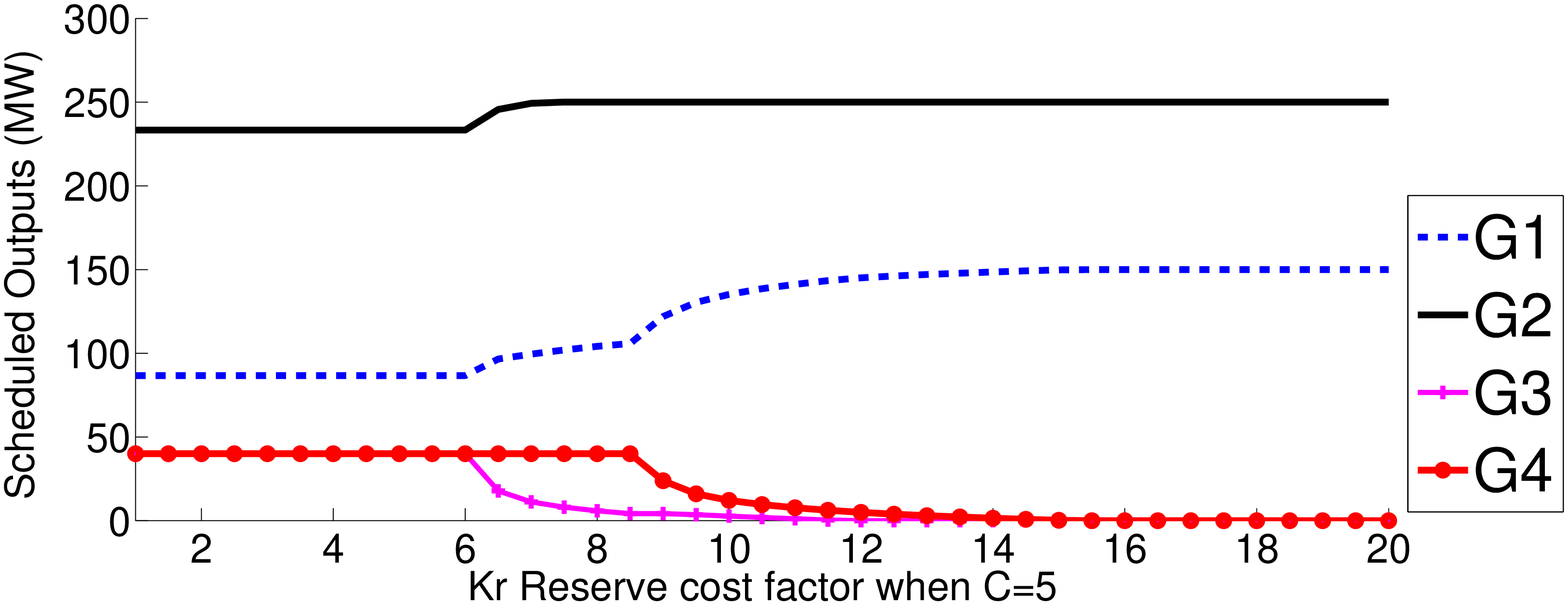}}
\subfloat[$c$=20]{\label{fig:C20kr_outs}\includegraphics[width=0.25\textwidth]{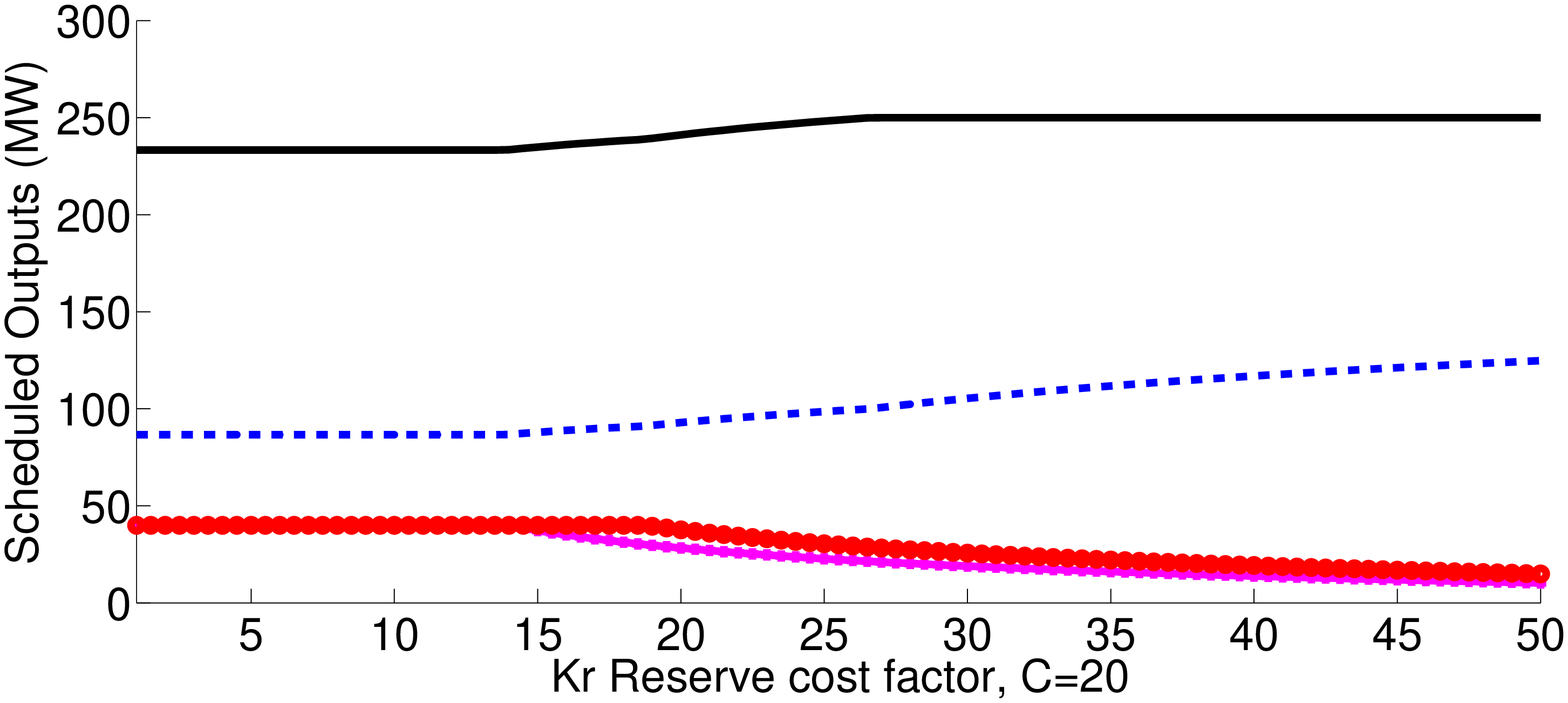}} \\
\hspace{0.3in}\parbox{3.4in}{\caption[Generators' outputs vs. reserve cost coefficient ($k_r$),]{Generator outputs vs. reserve cost coefficient ($k_r$) for two values of scale factor ($c$) in base case 400 $MW$}\label{fig:kroutsc}}
\end{figure}

 Fig. \ref{fig:kroutsc} shows the outputs of generators for a variation of reserve cost coefficient $k_r$ for two values of scale factor $c=5$ and $c=20$ in order to see where the critical change in wind power schedule begins.
  In Fig. \ref{fig:C5kr_outs} when $c$=5, the critical change in wind power schedule starts when $k_r=6$ for generator (3) and $k_r=8.6$ for generator (4). The drop of outputs happens in generators (3) before generator (4) because generator (3) has a higher direct cost (8 $\$/MW.hr$) than that of generator (4) (6 $\$/MW.hr$). While in the other case when c=20 as in Fig. \ref{fig:C20kr_outs}, the change of wind power scheduling occurs at higher values of $k_r$ because in this case, the scale factor $c$ of Weibull distribution of wind speed has a higher value.\\
\subsubsection{The Effects of Penalty Cost Coefficient}
\begin{figure}[!ht]
\centering
\subfloat[Generators' Outputs]{\label{fig:kpc}\includegraphics[width=0.25\textwidth]{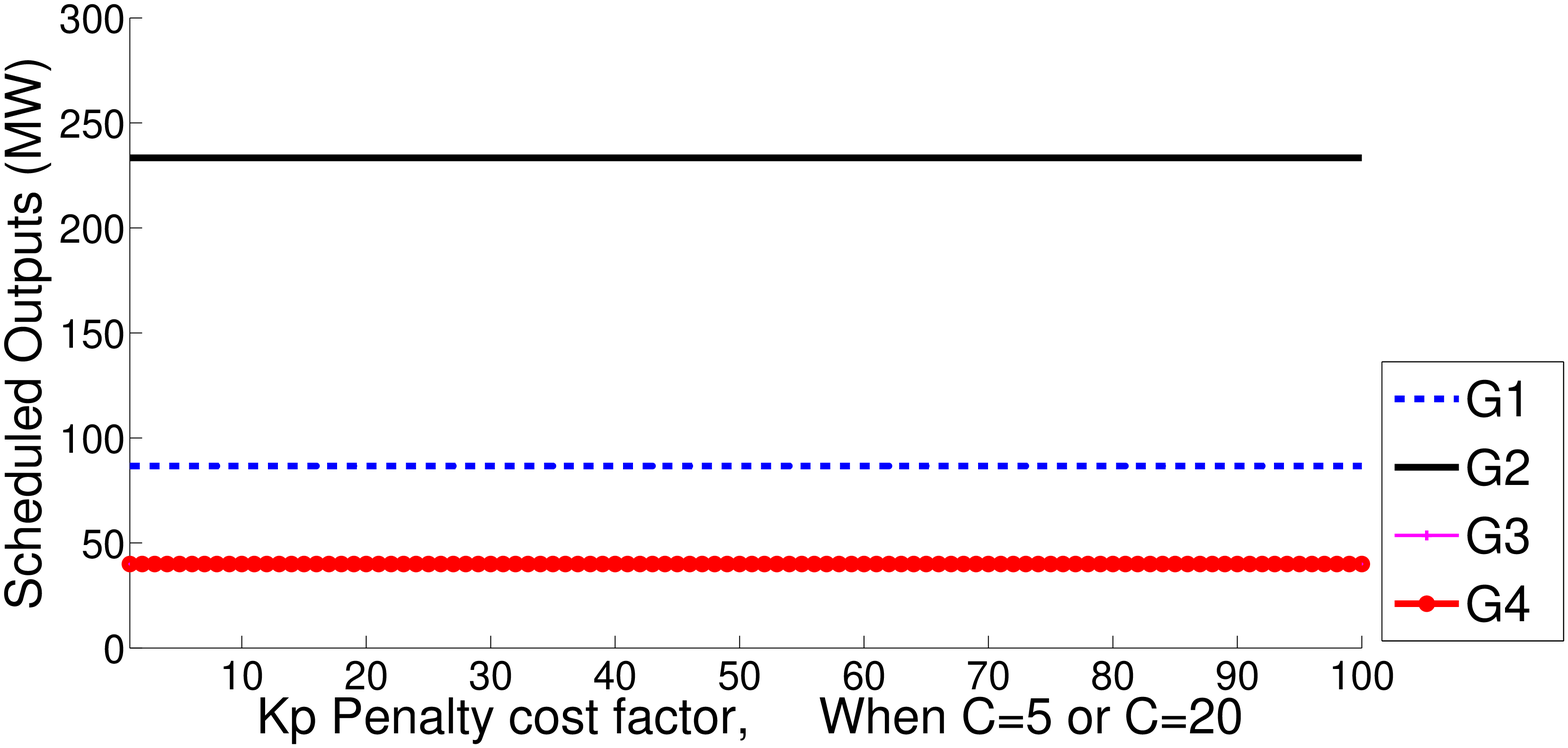}}
\subfloat[Penalty Cost $C_p$]{\label{fig:Cpc5_20}\includegraphics[width=0.25\textwidth]{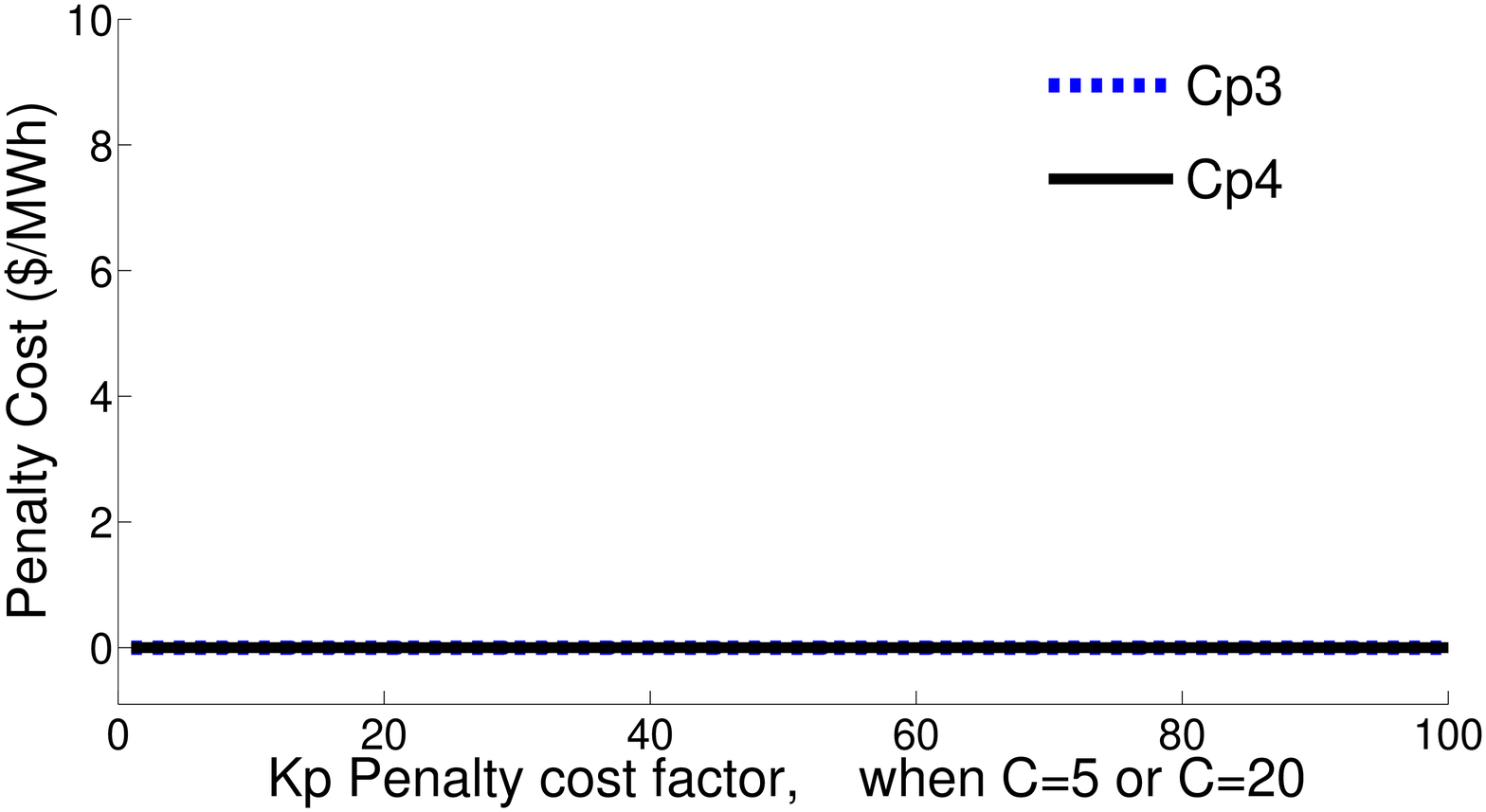}} \\
\hspace{0.3in}\parbox{3.4in}{\caption[Generators' outputs and penalty cost $C_p$ vs. penalty cost,]{Generators' outputs and penalty cost $C_p$ vs. penalty cost coefficient $k_p$ for two values of scale factor c in base case $400 MW$}\label{fig:wckpouts}}
\end{figure}

When $k_r=0$ and $k_p\neq0$, the schedule of generators as in Fig. \ref{fig:kpc} for various values of $k_p$ remains constant for different values of scale factor $c$.
In this case it should get all available wind power since there is a penalty cost for a surplus wind power. Fig. \ref{fig:Cpc5_20} shows that penalty cost $C_p$=0, thereby all available wind power is scheduled from both wind-powered generators as Fig, \ref{fig:kpc} shows.\\
\subsubsection{The Effects of The Reserve and Penalty Cost Coefficients}
\begin{figure}[!ht]
\centering
\subfloat[$k_r=0$]{\label{fig:kr0}\includegraphics[width=0.25\textwidth]{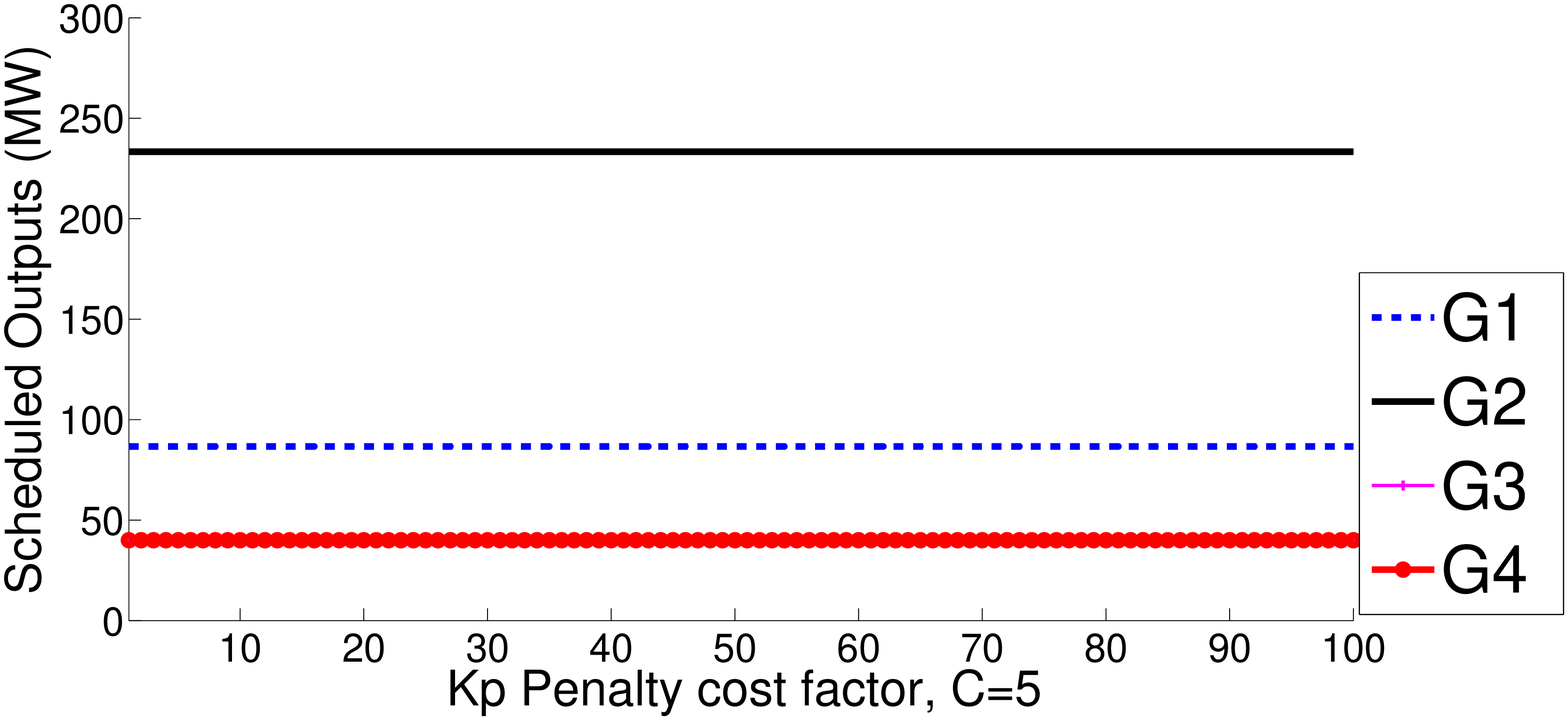}}
\subfloat[$k_r=60$]{\label{fig:kr60}\includegraphics[width=0.25\textwidth]{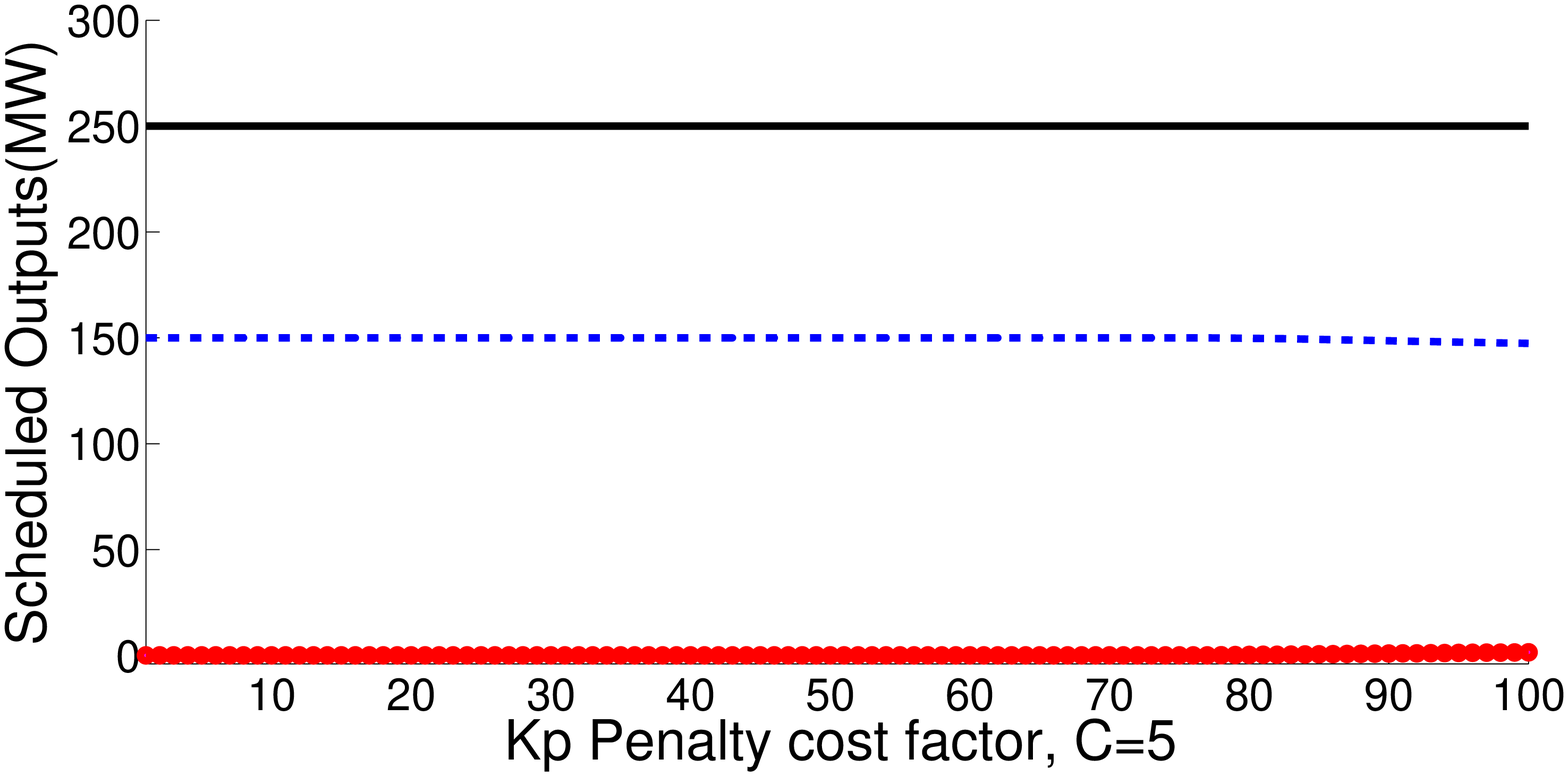}}
\hspace{0.3in}\parbox{3.4in}{\caption[Generators' outputs vs. penalty cost coefficient $k_p$,]{Generators' outputs vs. penalty cost coefficient $k_p$ for some values of reserve cost coefficient $k_r$ in the base case $400 MW$ and $c=5 m/s$}\label{fig:kr_kp_c}}
\end{figure}

 The effect of both of the reserve and the penalty cost coefficients of the economic dispatch outcomes) is illustrated in Fig. \ref{fig:kr_kp_c} for the base case with a scale factor $c$=5 and the wind cost coefficients are not equal to zero ($k_r\neq0$ and $k_p\neq0$).

When the reserve cost coefficient $k_r$ increases, the scheduled wind power decreases. Until no scheduled power comes from wind-powered units when $k_r\geq60$; because the high value of $k_r$ makes the wind power to be not an economic option. Thereby, all the scheduled power comes from thermal-powered units for any value of $k_p$, as in Fig. \ref{fig:kr60}.

\begin{figure}[!ht]
\centering
\subfloat[$c=10$]{\label{fig:kpkr20c10}\includegraphics[width=0.25\textwidth]{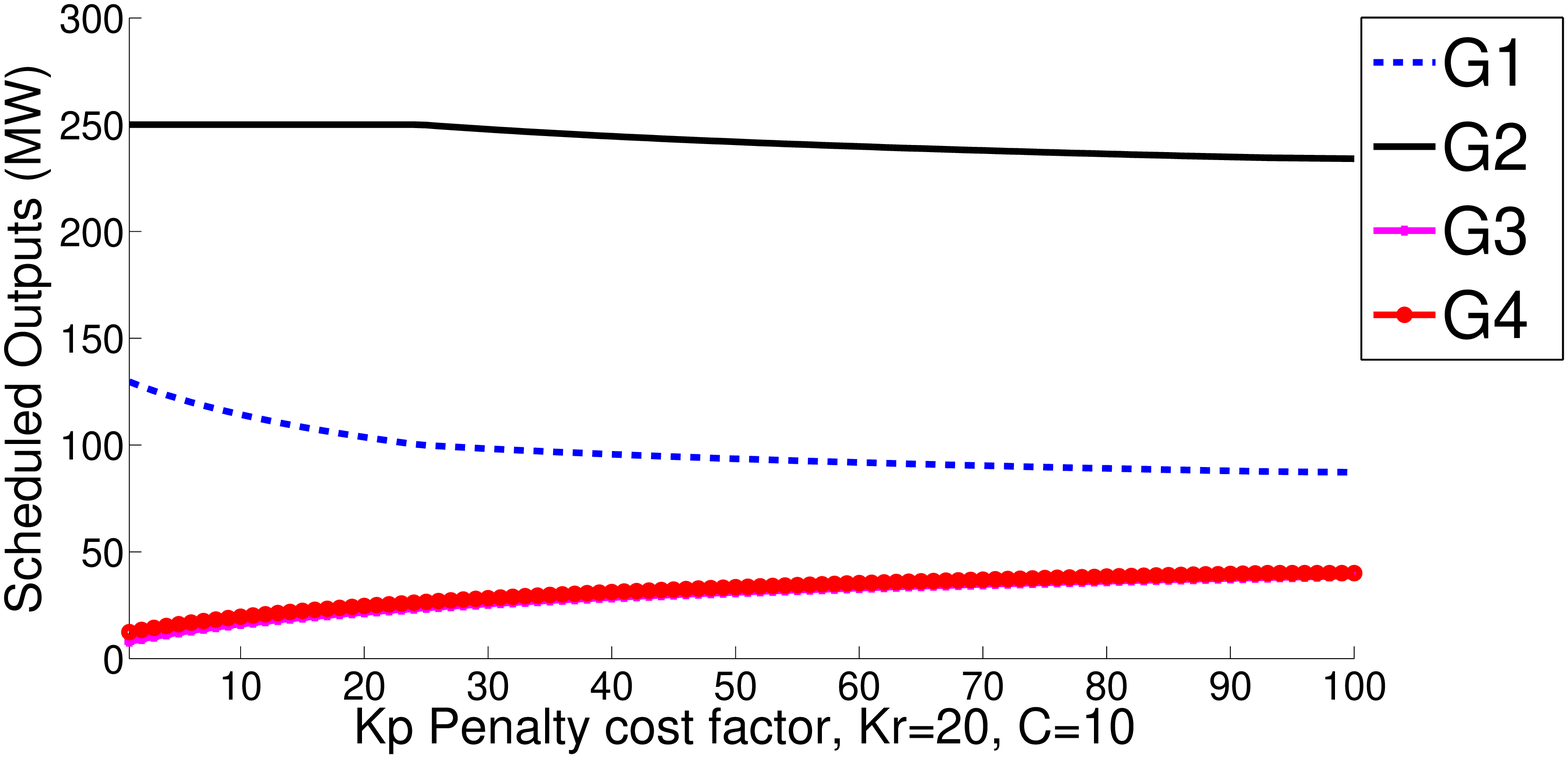}}
\subfloat[$c=20$]{\label{fig:kpkr20c20}\includegraphics[width=0.25\textwidth]{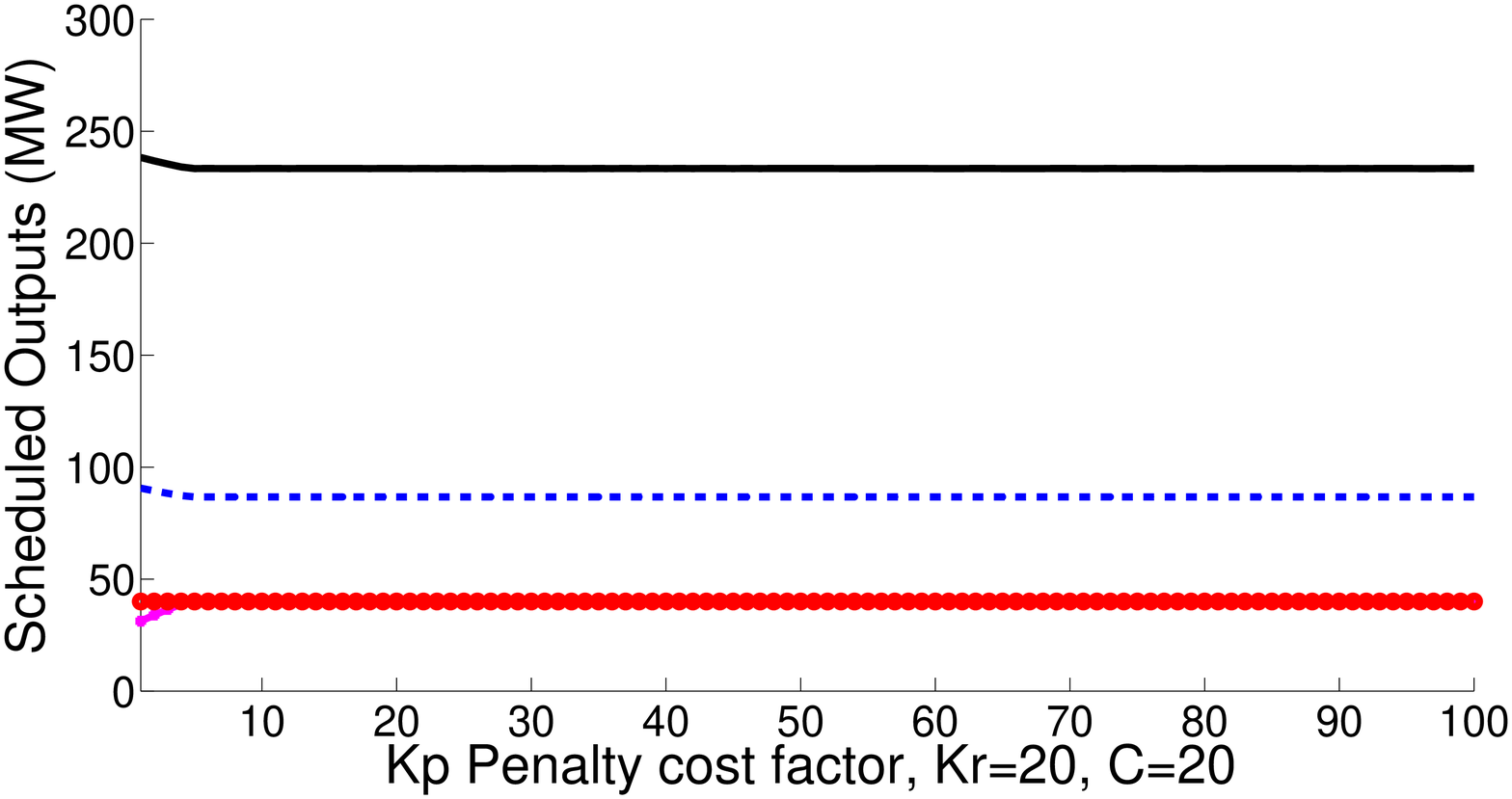}}
\hspace{0.3in}\parbox{3.4in}{\caption[Generators' outputs vs. penalty cost coefficient $k_p$ when $k_r$ =20,]{Generators' outputs vs. penalty cost coefficient $k_p$ for two values of scale factor $c$ when $k_r$=20 }
\label{fig:kpkr20c1020}}
\end{figure}

Fig. \ref{fig:kpkr20c1020} can be considered as a part of Fig. \ref{fig:kr_kp_c} when $kr$=20 but now for two higher values of the scale factor $c$=10 and $c$=20.

As it is shown in Fig. \ref{fig:kpkr20c1020}, with a higher scale factor $c$ of the probability distribution of wind speed, the outputs of wind-powered generators become higher as well. Furthermore, Fig. \ref{fig:kpkr20c10} illustrates that the wind power outputs will increase with higher values of the penalty cost coefficient $k_p$.\\
This is what happens when the utility does not own the wind turbine, therefore the scheduled wind power is produced as a compromise between the penalty cost and the reserve cost of wind power.


\section{Conclusion}  \label{conclusion}

The PSO algorithm is appropriate for finding the optimal economic dispatch of generators' outputs that include wind-powered generators.
Moreover, PSO algorithm is simple and easy to apply since it has fewer number of parameters to deal with comparing to other modern optimization algorithms. PSO can be applied in wind power bid marketing between electric power operators.
The used model of the economic dispatch that consideres the wind power uses the manipulation of the probability of the underestimation and overestimation of the availability of the wind power. It also takes into account whether the utility owns wind turbines or not; these are the main features of this model. 
The variations of wind speed parameters and their impacts on the total cost investigated by 6-bus system, some valuable conclusions have been noticed.
The incremental reserve and penalty costs of the available wind power can be compared to the incremental cost in the conventional-thermal units that have a quadratic cost; this comparison could lead to useful simplifications of the economic dispatch models that include thermal and wind power.
 PSO algorithm needs some work on selecting proper parameters and it also needs some further mathematical description for its convergence. 


\bibliographystyle{ieeetr}
\bibliography{Refs1}

\begin{thebibliography}{10}

\bibitem{Wang2005}
C.-R. Wang, H.-J. Yuan, Z.-Q. Huang, J.-W. Zhang, and C.-J. Sun, ``A modified
  particle swarm optimization algorithm and its application in optimal power
  flow problem,'' in {\em Proc. Int Machine Learning and Cybernetics Conf},
  vol.~5, pp.~2885--2889, 2005.

\bibitem{Vlachogiannis2009}
J.~G. Vlachogiannis and K.~Y. Lee, ``Economic load dispatch---a comparative
  study on heuristic optimization techniques with an improved coordinated
  aggregation-based pso,'' vol.~24, no.~2, pp.~991--1001, 2009.

\bibitem{hwary_enhanced}
M.~R. AlRashidi, M.~F. AlHajri, and M.~E. El-Hawary, ``Enhanced particle swarm
  optimization approach for solving the non-convex optimal power flow,'' {\em
  World Academy of Science, Engineering and Technology}, vol.~62, pp.~651--655,
  2010.

\bibitem{Ken}
J.~Kennedy and R.~Eberhart, ``Particle swarm optimization,'' in {\em Proc.
  Conf. IEEE Int Neural Networks}, vol.~4, pp.~1942--1948, 1995.

\bibitem{Review2009}
R.~Boqiang and J.~Chuanwen, ``A review on the economic dispatch and risk
  management considering wind power in the power market,'' {\em Renewable and
  Sustainable Energy Reviews}, vol.~13, no.~8, pp.~2169--2174, 2009.

\bibitem{Hetzer}
J.~Hetzer, D.~C. Yu, and K.~Bhattarai, ``An economic dispatch model
  incorporating wind power,'' vol.~23, no.~2, pp.~603--611, 2008.

\bibitem{Wood}
A.~J. Wood and B.~F. Wollenberg, {\em Power Generation, Operation and Control}.
\newblock New York: Wiley, 2nd~ed., 1996.

\bibitem{Wang1}
L.~Wang and C.~Singh, ``Pso-based multi-criteria economic dispatch considering
  wind power penetration subject to dispatcher's attitude,'' in {\em Proc. 38th
  North American Power Symp. NAPS 2006}, pp.~269--276, 2006.

\bibitem{wind&solar}
M.~R. Patel, {\em Wind and Solar Power Systems}.
\newblock CRC Press, 1999.

\bibitem{Gaing}
Z.-L. Gaing, ``Particle swarm optimization to solving the economic dispatch
  considering the generator constraints,'' vol.~18, no.~3, pp.~1187--1195,
  2003.

\bibitem{abuella}
M.~A. Abuella, ``Study of particle swarm for optimal power flow in ieee
  benchmark systems including wind power generators.'' [Online]. Available:
  {http://opensiuc.lib.siu.eud/theses}, 2012.

\end{thebibliography}
\

\section*{biographies}

\footnotesize{\textbf{Mohamed A. Abuella} (IEEE student member). He is a PhD student at University of North Carolina at Charlotte. He received his Bachelor's degree from College of industrial Technology, Misurata, Libya in 2008, and MSc degree from Southern Illinois University Carbondale in 2012. His research interests are in planning and operations of power systems.}\\

\footnotesize{\textbf{C.J. Hatziadoniu} (IEEE M'87, SM'06). He is a professor of electrical and computer engineering at Southern Illinois University Carbondale. His research interests include power systems control and protection and application of power electronics to power systems.}

%
%
%

\end{document}